# KEEPERS OF THE DOUBLE STARS

## Joseph S. Tenn

*Sonoma State University, Rohnert Park, CA 94928, USA.*
E-mail: joe.tenn@sonoma.edu

**Abstract:** Astronomers have long tracked double stars in efforts to find those that are gravitationally-bound binaries and then to determine their orbits. Early catalogues by the Herschels, Struves, and others began with their own discoveries. In 1906 court reporter and amateur astronomer Sherburne Wesley Burnham published a massive double star catalogue containing data from many observers on more than 13,000 systems. Lick Observatory astronomer Robert Grant Aitken produced a much larger catalogue in 1932 and coordinated with Robert Innes of Johannesburg, who catalogued the southern systems. Aitken maintained and expanded Burnham's records of observations on handwritten file cards, and eventually turned them over to the Lick Observatory, where astrometrist Hamilton Jeffers further expanded the collection and put all the observations on punched cards. With the aid of Frances M. "Rete" Greeby he made two catalogues: an Index Catalogue with basic data about each star, and a complete catalogue of observations, with one observation per punched card. He enlisted Willem van den Bos of Johannesburg to add southern stars, and together they published the *Index Catalogue of Visual Double Stars, 1961.0*. As Jeffers approached retirement he became greatly concerned about the disposition of the catalogues. He wanted to be replaced by another "double star man," but Lick Director Albert E. Whitford had the new 120-inch reflector, the world's second largest telescope, and he wanted to pursue modern astrophysics instead. Jeffers was vociferously opposed to turning over the card files to another institution, and especially against their coming under the control of Kaj Strand of the United States Naval Observatory. In the end the USNO got the files and has maintained the records ever since, first under Charles Worley, and, since 1997, under Brian Mason. Now called the Washington Double Star Catalog (WDS), it is completely online and currently contains more than 1,200,000 measures of more than 125,000 star systems.

**Keywords:** double stars, binary stars, William Herschel, John Herschel, James South, Wilhelm Struve, Otto W. Struve, Sherburne W. Burnham, Robert Aitken, Robert Innes, Hamilton Jeffers, Willem van den Bos, Frances Greeby, Albert Whitford, Kaj Strand, Charles Worley, Brian Mason, William Hartkopf, Pulkova, Lick Observatory, United States Naval Observatory.

## 1 INTRODUCTION

*Double stars*—pairs of stars that appear close together in the sky—have been known since ancient times. Ptolemy described the pair now known as $v^1$ and $v^2$ Sgr as a double star nearly two thousand years ago. Mizar (Figure 1) was observed as double through a telescope in 1617 by Galileo's student Benedetto Castelli (1578–1643) (Ondra, 1999; Fedele, 1949).

In 1767 the English clergyman John Michell (Figure 2), who would later predict black holes, applied his statistical skill to the by-then substantial number of double stars known:

> We may from hence, therefore with the highest probability conclude (the odds against the contrary opinion being many million millions to one) that the stars are really collected together in clusters in some places, where they form a kind of systems, whilst in others there are either few or none of them, to whatever cause this may be owing, whether to their mutual gravitation, or to some other law or appointment of the Creater [sic]. And the natural conclusion from hence is that it is highly probable in particular, and next to a certainty in general, that such double stars, &c. as appear to consist of two or more stars placed very near together, do really consist of stars placed near together, and under the influence of some general law, whenever the probability is very great, that there would not have been any such stars so near together, if all those, that are not less bright than themselves, had been scattered at random through the whole heavens. (Michell, 1767: 249–250).

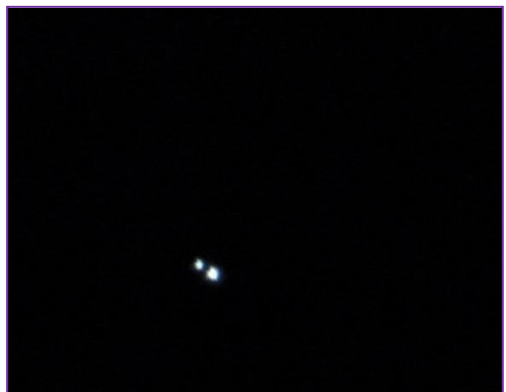

Figure 1: Mizar, the first double star seen through a telescope (courtesy: George Kristiansen).

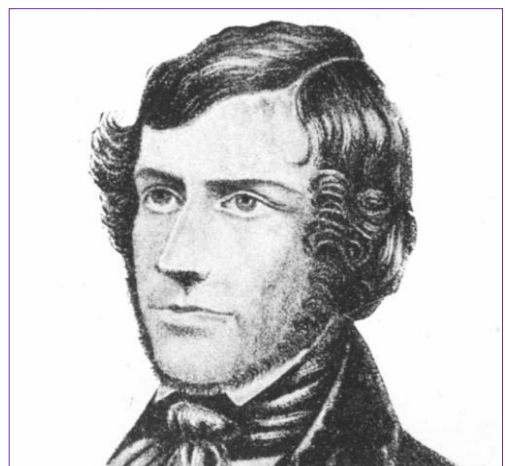

Figure 2: John Michell (1724–1793) (http://wdict. net/word/john+michell/).

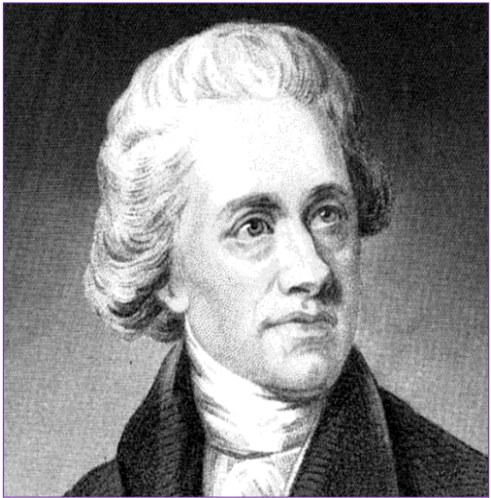

Figure 3: William Herschel (1738–1822), the founder of double star astronomy (after Holden, 1881: Frontispiece).

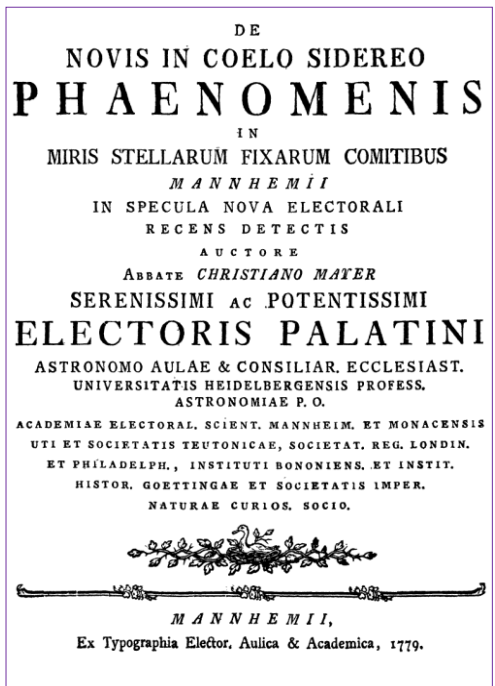

Figure 4: Title page of first double star catalogue (after Mayer, 1779).

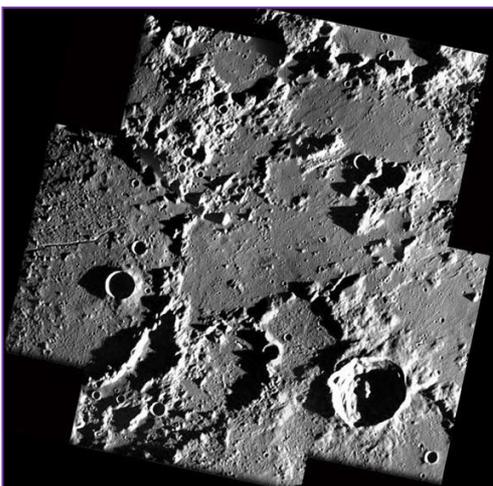

Figure 5: The lunar crater C. Mayer is at lower right (ESA/Space-X Space Exploration Institute).

In 1803 William Herschel (Figure 3) proved that Michell was right in suspecting "… their mutual gravitation." After 25 years of recording the positions of α Geminorum, γ Leonis, ε Bootis, ζ Herculis, δ Serpentis and γ Virginis (Herschel, 1803), he had become convinced that all were orbiting systems, or *binary stars*, a term he had coined the previous year (Herschel, 1802: 480-481). He had started these observations confident that the pairs were chance optical alignments, and he had measured them hoping to be able to detect stellar parallax by measuring shifts in the positions of the brighter, presumably nearer, stars with respect to the dimmer, more distant ones.

By this time astronomers were adept at applying Newtonian physics to such systems. If distances could be determined, as they could starting in the late 1830s with the first parallax measurements by Friedrich Wilhelm Bessel (1784–1846), Friedrich Georg Wilhelm Struve (1793–1864), and Thomas James Henderson (1798–1844), then angular separations in arcseconds could be converted to separation distances in astronomical units. Then the separation distance (in au) cubed divided by the period (in years) squared gave the sum of the masses of the two stars (in solar masses). Individual stellar masses could even be computed if the relative distances from the stars to the center of mass of the system could be found.[1] To this day this is the only direct means to determine stellar masses. Binary stars are important!

## 2 EARLY CATALOGUES

But which pairs of double stars are binary? How common are double stars? How can an astronomer who discovers a close pair learn whether there are earlier observations which could be combined with his own to determine whether the system is binary? The solution to all of these problems is to construct catalogues.

The first catalogue of double stars was published by Christian Mayer (1719–1783) in 1779. This multitalented and multilingual Jesuit priest, born in what is now the Czech Republic, listed 72 double star systems in a small book (Figure 4; Mayer, 1779; Schlimmer, 2007). Most of these were discovered by him, although some had been observed by others, such as William Herschel. He presented separation angles, but neither in this book nor in a slightly longer list published later in the *Astronomische Jahrbuch* for 1784 (Mayer, 1781) did he give position angles. He merely stated that the dimmer star was southwest or south, etc. of the brighter one. There do not appear to be any portraits of Mayer, who taught physics and mathematics at Heidelberg and served as court astronomer in Mannheim until the Jesuit order was dissolved by the Pope in 1773, but he can be represented by the fine lunar crater that carries his name (Figure 5).

A number of others published catalogues of double stars over the next hundred years. Many consisted of simple lists of doubles discovered or rediscovered by the author, but some included observations by others. The most important catalogues were by two fathers and their sons. William Herschel published a list of 269 doubles, 227 of which he claimed to be the first to see, in 1782 (Herschel, 1782), the year he took up his few duties as astronomer to England's King George III. Three years later he followed up with a list of 434 more, all of his own discovery. As he put it,

> The happy opportunity of giving all my time to the pursuit of astronomy, which it has pleased the Royal Patron of this Society to furnish me with, has put it in my power to make the present

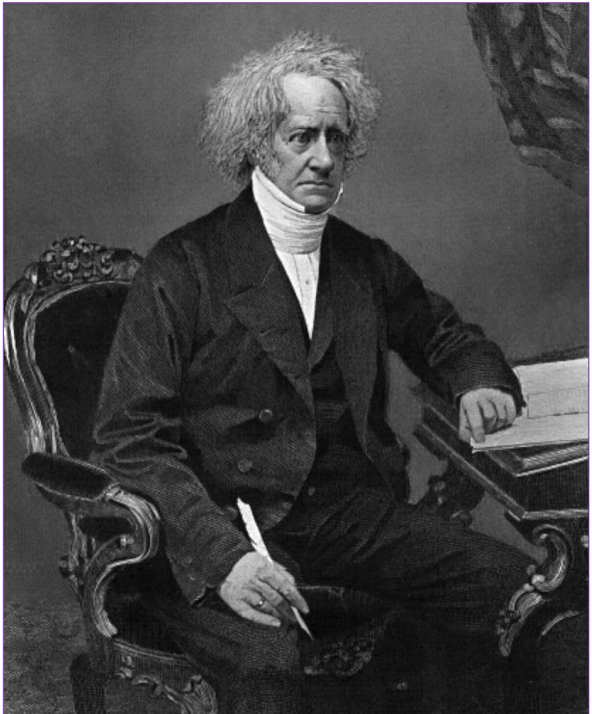

Figure 6: John Herschel (1792–1871) (after Duyckinick, 1873).

> collection much more perfect than the former; almost every double star in it having the distance and position of its two stars measured by proper micrometers; and the observations have been much oftener repeated. (Herschel, 1785: 40).

His last, less detailed, catalogue was presented when he was in his eighties and serving as first President of what would become the Royal Astronomical Society (Herschel, 1821). Shortly afterward his son, John (Figure 6), published, with surgeon and wealthy amateur James South (Figure 7), his first catalogue, with 380 stars arranged in order of right ascension and including accounts of other measurements besides their own (Herschel and South, 1824).

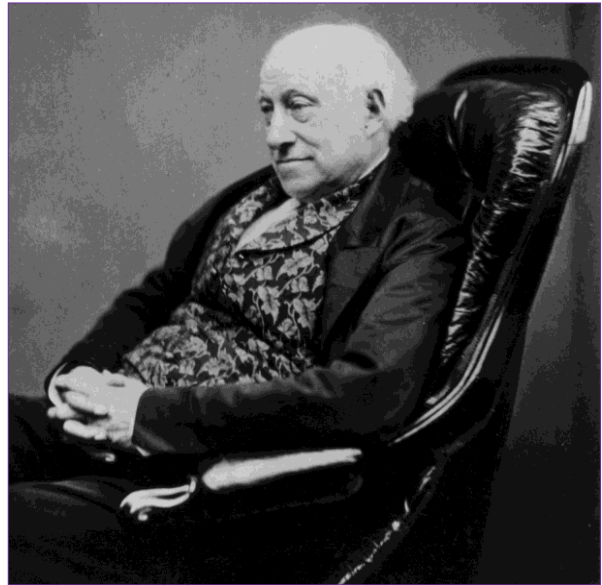

Figure 7: James South (1785–1867) (NPG P120(20) by Maull & Polyblank (detail), 1855 © National Portrait Gallery, London, used by permission).

Meanwhile in Dorpat, Russia (now Tartu, Estonia), Wilhelm Struve (Figure 8) was publishing a catalogue of 795 stars in 1822 (Struve, 1822) and a much bigger one, with 3112 stars, five years later (Struve, 1827). This catalogue (Figure 9), based on Struve's examination with the Dorpat Observatory's new 9.6-in Fraunhofer refractor of all stars (~120,000) brighter than $9^{th}$ magnitude and north of −15°, included more than 2500 newly-discovered doubles. It became the world's most important catalogue for some time. It was renowned for the "… extraordinary accuracy for the epoch …" (Jackson, 1922: 4) of the author's measurements despite the fact that he sometimes ex-amined as many as 400 stars per hour, finding one out of every thirty-five to have a companion within 32 arcseconds (Batten, 1988: 51). Struve published additional catalogues (Struve, 1837;

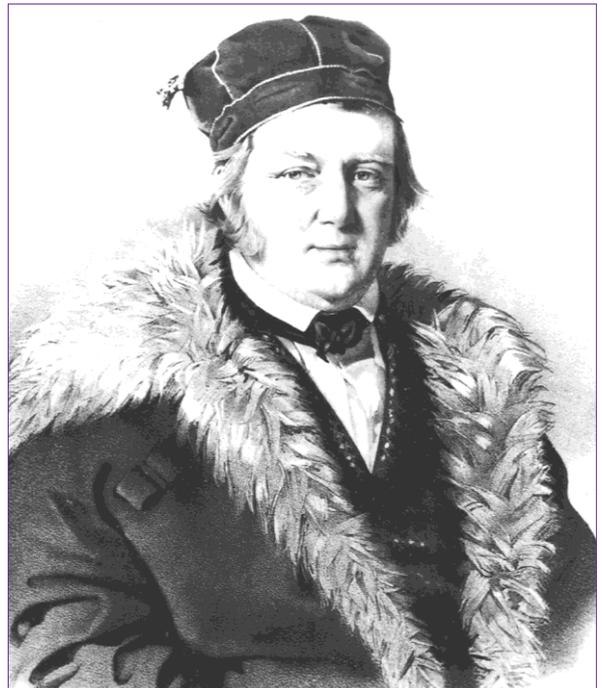

Figure 8: Wilhelm Struve (1793–1864) (after a lithograph made from a painting by C.A. Jensen, 1843).

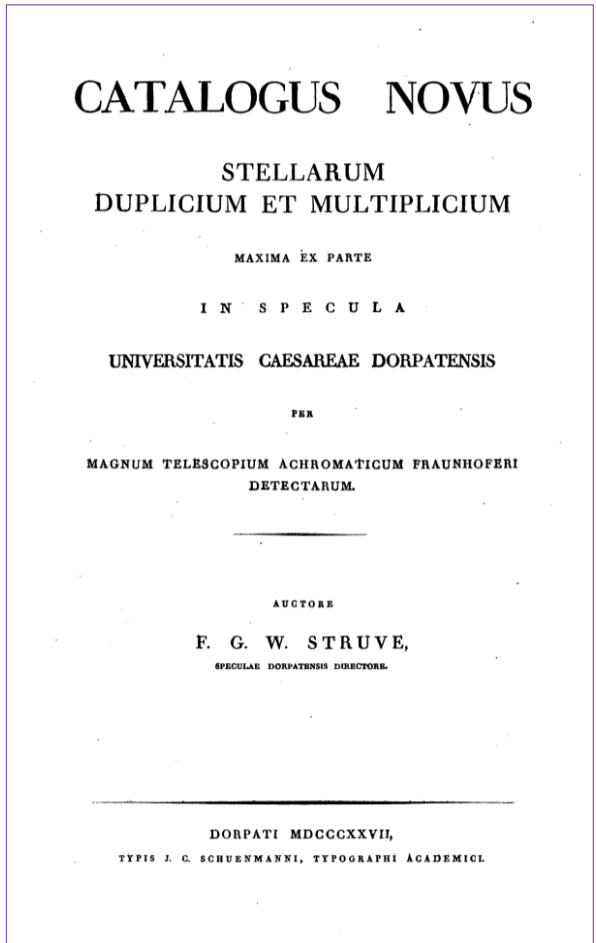

Figure 9: The title page of F.G.W. Struve, 1827 (Google Books).

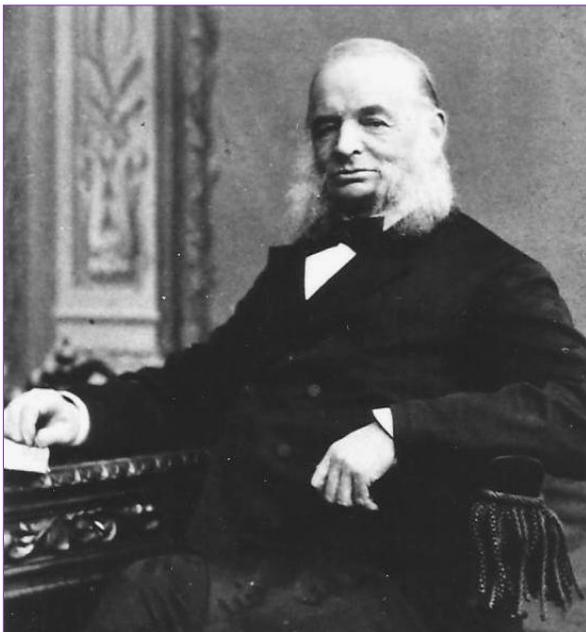

Figure 10: Otto W. Struve (1819–1905) (photograph by W. Clasen).

1852) in later years after becoming founding Director of the Pulkovo Observatory outside St. Petersburg, where he secured a 15-in refractor.

Struve's son, Otto W. (Figure 10), who worked with his father at Pulkovo from 1839 and would succeed him as Director in 1862, added another major catalogue in 1843 (Struve, O.W., 1843) and supplemented it with several much smaller lists until 1878, giving him a total of 547 systems catalogued.

By the late nineteenth century, many astronomers thought that most binaries that could be discovered had been found. For example, according to Robert Aitken in 1935:

> The feeling that the Herschels, South, and the Struves had practically exhausted the field of double star discovery, at least for astronomers in the northern hemisphere, continued for thirty years after the appearance of the Pulkowa Catalogue in 1843. (Aitken, 1964: 20).

There was, of course, ample reason to re-observe the known doubles. John Herschel, who published a number of lists of newly-discovered double stars in the 1820s and 1830s, was working on a very large catalogue of all known pairs when he died in 1871. It was completed by others and published posthumously (Herschel, 1874) and contained about 10,300 double star systems. However, Herschel had only compiled the measurements of separations and orientations for two-fifths of the systems when he died. His editors published just the list of the stars with their co-ordinates and precessions. It was in this catalogue that John Herschel introduced a set of symbols that was to last a long time: stars in Wilhelm Struve's final catalogue are denoted by the Greek letter $\Sigma$, while those in Otto Struve's catalogues are labeled O$\Sigma$ and O$\Sigma\Sigma$. William Herschel's discoveries are labeled H, and John Herschel's, h. In each case the letters are followed by catalogue numbers. There were other letters as well, and later observers would expand this system with, for example, $\beta$ for Burnham. In a modified form, with STF replacing $\Sigma$ for Wilhelm Struve and STT replacing O$\Sigma$ for Otto Struve, it is retained today as a cross reference in the Washington Double Star Catalog (WDS, n.d.).

## 3 BURNHAM AND THE BDS

The American Sherburne Wesley Burnham (Figure 11) raised double star cataloguing to new heights. He had little formal education but taught himself shorthand and became a stenographer, a profession of high standing and pay in his time. After recording courts martial for the Union Army in the American Civil War, he became a court reporter in Chicago and bought a six-inch (15-cm) Clark refractor, which he used in his spare time to discover hundreds of double stars that had been missed by professional astronomers with larger instruments. Soon he was elected a Fellow of the Royal Astronomical Society and allowed to use the 18.5-inch (47-cm) refractor (Figure 12) of the Dearborn Observatory[2] to look for more doubles. He gave up his amateur standing to serve on the initial staff of the Lick Observatory from 1888 to 1892, but resigned to go back to Chicago and court reporting. The needs of his large family for schooling, which was not available on Mount Hamilton, the opportunity to double his salary, and his (widely shared) dislike of Lick Director Edward Singleton Holden (1846–1914) were ample reasons for the move (Osterbrock, et al., 1988). For the rest of his career Burnham worked in the federal courts of Chicago, but his renown among astronomers enabled him to make avocational use of the long refractors of Yerkes Observatory, the Washburn Observatory, and the United States Naval Observatory (USNO), as well as the Dearborn. By the end of the century he had ure 13), the Director of the Flower (later Flower and Cook) Observatory of the University of Pennsylvania.

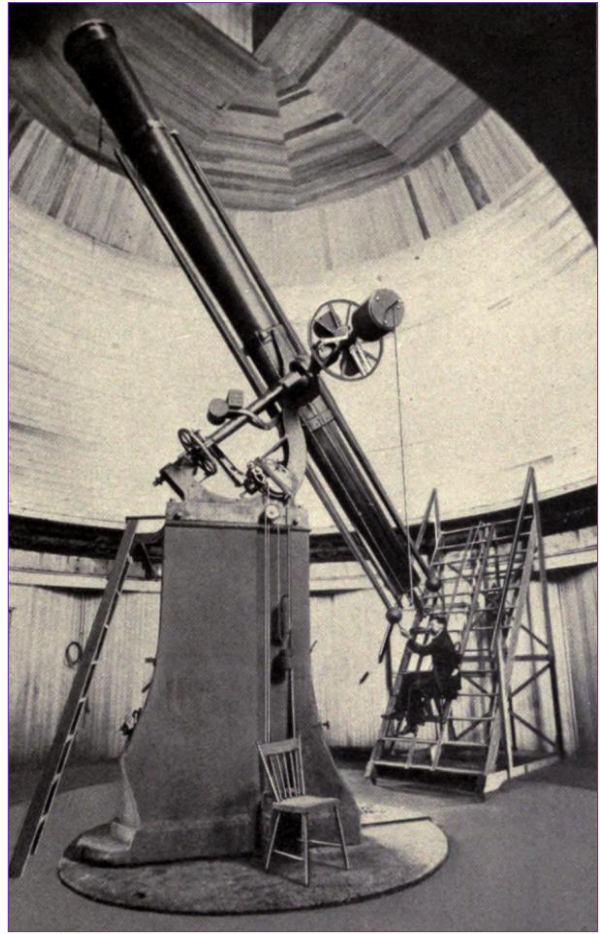

Figure 12: The 18½-in Clark Refractor of the Dearborn Observatory (after Burnham, 1900: opposite page xii).

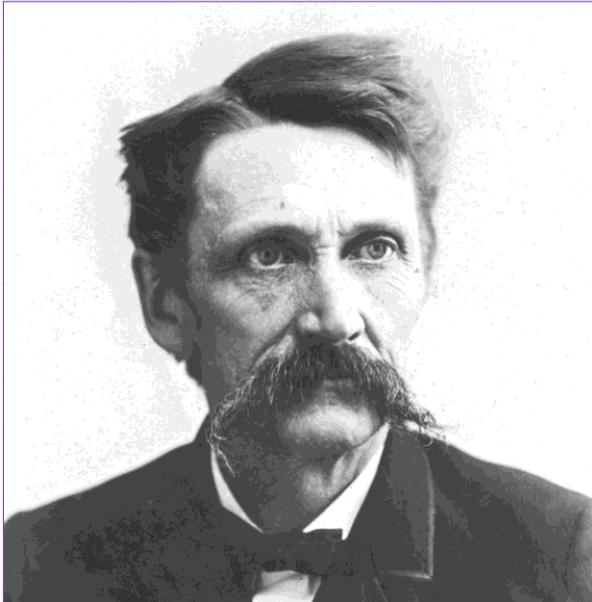

Figure 11: Sherburne Wesley Burnham (1838–1921) (University of Chicago Photographic Archive, apf6-02814, Special Collections Research Center, University of Chicago Library).

discovered and published some 1290 pairs of his own (Burnham, 1900), and he had been collecting observations of others for many years.

In 1906 Burnham published his magnum opus, *A General Catalogue of Double Stars within 121° of the North Pole* (Burnham, 1906), containing observations of 13,665 star systems from around the world. Known as the BDS (Burnham Double Stars), it would serve astronomers for decades. Immediately after completing this work, he started preparing for a second edition by making a handwritten card catalogue of all subsequent observations with one card for each observation. When, he found himself unable to continue, in about 1912, he turned the cards over to Eric Doolittle (Fig-

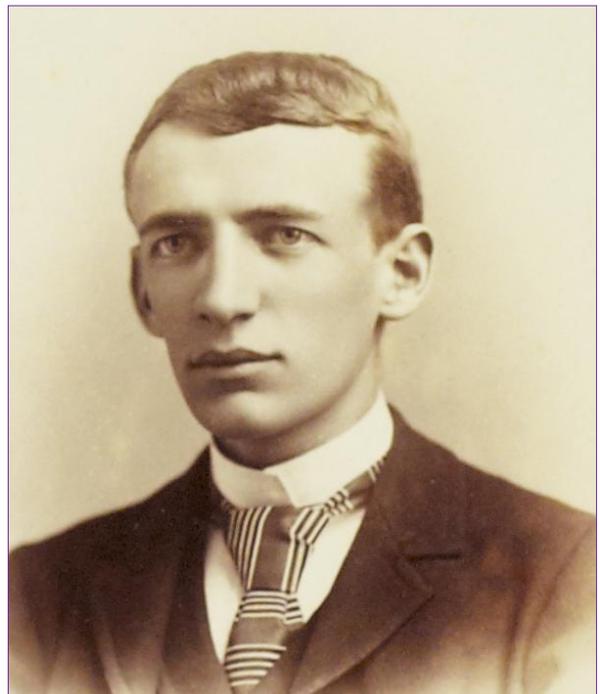

Figure 13: Eric Doolittle (1870–1920) as a student at Lehigh University in 1891 (courtesy: Special Collections, Lehigh University Libraries, Bethlehem, Pennsylvania).

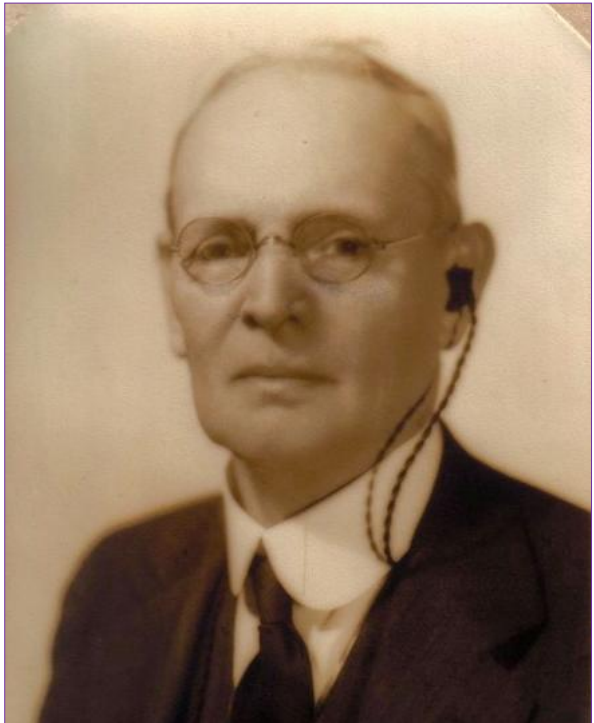

Figure 14: Robert Grant Aitken (1864–1951) (courtesy of his grandson, Robert Aitken).

## 4  AITKEN AND INNES, THE ADS AND THE SDS

Doolittle had maintained the card catalogue for only a few years when his health began to fail, so he prevailed on Lick Observatory astronomer Robert Aitken (Figure 14) to take over the responsibil-ity. Aitken began maintaining the card catalogue shortly after Doolittle's death in 1920.

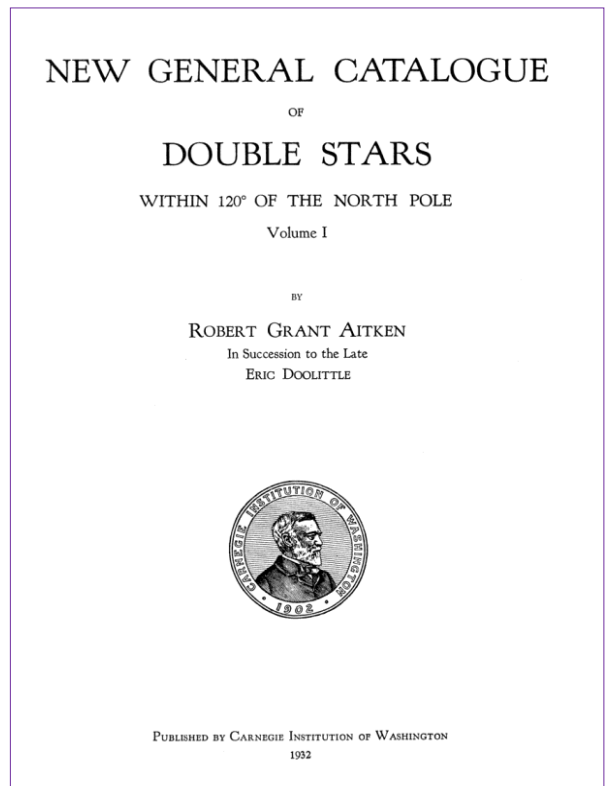

Figure 15: The title page of the ADS (Aitken, 1932).

Born in a mining town in California shortly after the California gold rush, Robert Aitken crossed the continent to study for the ministry at Williams College in Massachusetts. There he came under the influence of astronomy professor Truman H. Safford (1836–1901) and earned his B.A. in mathematics instead. Shortly after graduation he taught both subjects at the University of the Pacific, then in San Jose at the foot of Mount Hamilton, where the Lick Observatory had gone into operation just six years earlier. In 1894 Aitken wrote Holden, asking to visit Lick during the summer in order to learn to use the transit and micrometer in his teaching. He was successful, and soon he was asking to study astronomy full-time if a way could be found to support his family. He was eager to leave teaching and become a research astronomer (Tenn, 1993; 2013a).

In 1895, Aitken, his wife, and three young sons took the horse-drawn stage up Mount Hamilton, initially so that he could be a summer student. Offered a one-year appointment after just two weeks at Lick, he stayed for forty years, ultimately becoming the Observatory Director.

From the beginning Aitken concentrated on visually measuring double stars with the 36-in and 12-in refractors. At first he measured stars on lists sent to him by Burnham, but soon he was discovering his own. In 1898 he and colleague William Joseph Hussey (1862–1926) divided the sky and set out to examine all stars down to the ninth magnitude and catalogue the doubles. Burnham had to add a supplement to the BDS to include nearly a thousand new systems discovered by Hussey and Aitken in just a few years. Most of the time Hussey used the big Lick Refractor and Aitken the 12-in. After Hussey left in 1905 to head up the new astronomy program at the University of Michigan, Aitken carried on alone. Now he had a great deal of time on the 36-in. Ultimately, Aitken discovered 3087 new double stars and made 26,650 double star observations (Couteau, 1988). After maintaining and expanding Burnham's card catalogue through the beginning of 1927, he set about producing his two-volume, *New General Catalogue of Double Stars within 120° of the North Pole* (Aitken, 1932, Figure 15). Both the title and the lengthy historical introduction make it clear that the book was to be considered a successor to BDS. And it was.

The ADS (Aitken Double Stars), as it became known, contained about 17,000 stars. Like Burnham before him, Aitken had had to find a patron to pay for the costly printing. Burnham had relied on philanthropist Catherine Wolfe Bruce[3] for his 1900 catalogue, while the Carnegie Institution of Washington published both his BDS and Aitken's ADS.

The ADS succeeded the BDS as the double star astronomers' chief reference work, except for stars far to the south. However, this time there was something new. Aitken had coordinated with Robert T.A. Innes (Figure 16), who was simultaneously preparing a double star catalogue of the southern sky. Born in Scotland, where he started publishing papers in celestial mechanics as a highly-skilled amateur despite leaving school at age twelve, Innes had emigrated to Australia, where he ran a liquor and wine business. There he became a serious amateur observer, publishing a short list of his own discoveries of double stars as early as 1894 and attempting unsuccessfully to publish a more general catalogue the following year (Orchiston, 2001; Astronomical Society of Southern Africa, n.d.). He moved to South Africa in 1896 for the opportunity to work in a professional observatory. By 1903 he was founding Director of what would later be called the Union Observatory, and near the end of his career he brought out what was intended to be a preliminary catalogue of mimeographed loose sheets (Innes, et al., 1927). Called the SDS (Southern Double Stars), it was in the same format as the ADS, and between them they covered the sky with some overlap.

Aitken eventually reported some 26,650 observations of double stars. He also spent a great deal of time on writing, speaking, and education and public outreach, much of it done through the Astronomical Society of the Pacific. He edited the *Publications of the ASP* from 1898 to 1942 and twice served as ASP President. He served as Associate Director of Lick from 1923 to 1930, under the thumb of Director William Wallace Campbell (1862–1938), who was in Berkeley, serving as President of the University of California, and finally as Director from 1930 to 1935. Aitken then retired and moved to Berkeley. He gave up maintaining the card catalogue and presented it, along with his personal library (some of it inherited from Burnham via Doolittle), to the Lick Observatory in 1944, when he was eighty. It was promptly turned over to Hamilton Jeffers.

## 5 JEFFERS, VAN DEN BOS, AND THE IDS

Hamilton Jeffers (Figure 17) was born in Pennsylvania, the son of a Professor of Biblical History, and he remained deeply conservative, both personally and politically, throughout his life. His older brother, Robinson (1887–1962), became a rather prominent poet and settled fairly near the Lick Observatory at Carmel on the central California coast. Jeffers first worked at Lick while a graduate student at the University of California from 1917 to 1921. He returned when Aitken hired him to work at Lick in 1924. He would remain there the rest of his career. In his early years he became an expert with the meridian circle, measuring positions of reference stars, and also with timekeeping and short wave radio. A man of many hobbies, he delighted in flying his own plane across the country to attend astronomical meetings. He was the principal user of the venerable 36-in Lick refractor both before and after service in WWII, and it was natural that he would succeed Aitken in maintaining the catalogue of double stars (Tenn, 2013b).

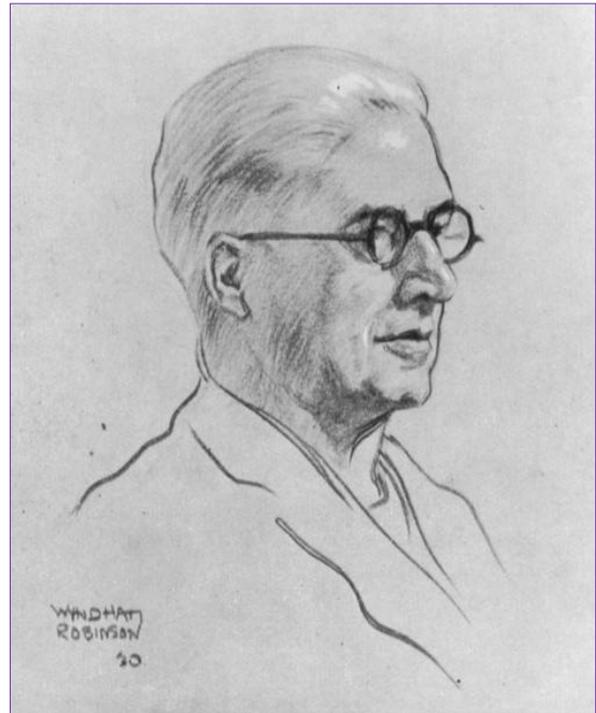

Figure 16: Robert Thorburn Ayton Innes (1861–1933) (after *Journal of the Astronomical Society of South Africa*, 3: 1).

Using the eyepiece interferometer invented by South African astronomer W.S. Finsen (1905–1979), Jeffers added greatly to Aitken's collection of file cards, which included data on about 17,000 double stars when he acquired it. By the early 1950s it was clear to Jeffers that the time had come to replace the handwritten cards of Burnham, Doolittle, and Aitken with new technology:

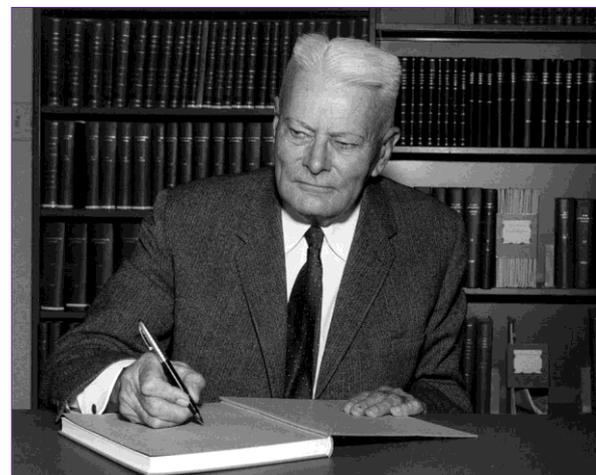

Figure 17: Hamilton Jeffers (1893–1976) (courtesy: Special Collections & Archives, University of California Santa Cruz).

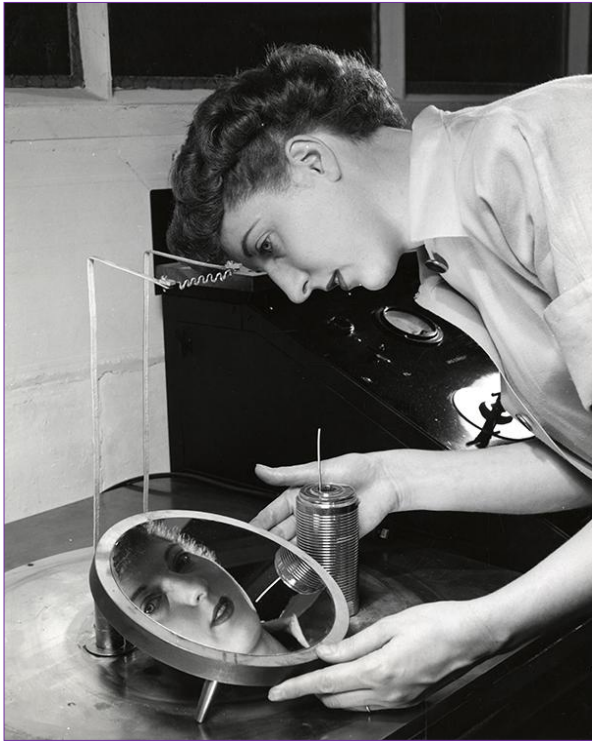

Figure 18: Frances M. 'Rete' Greeby (1921–2002) (courtesy: California Academy of Sciences).

machine readable punched cards. He secured a grant from the National Scientific Foundation and hired Frances M. 'Rete' Greeby to do the key-punching. Greeby had recently moved to Mt. Hamilton when her husband took a job at Lick; she had previously worked for the California Academy of Sciences in San Francisco, where she made the star plates for the Morrison Planetarium, as shown in Figure 18.

As he extended the punched card catalogue back to 1927.0, the date of the ADS, Jeffers realiz-ed that it would now be possible to combine all observations of double stars, north and south, into one catalogue. He joined forces with Willem van den Bos, the Dutch-born Johannesburg astronomer who as a young man had been a coauthor of the SDS and was now Director of the Union Observatory. So after punching 95,000 cards representing observations of double stars north of −20°, Greeby punched another 50,000 using data on southern stars sent by van den Bos, who became a close friend of Jeffers and visited Lick from 1961 to 1963 (Figure 19).

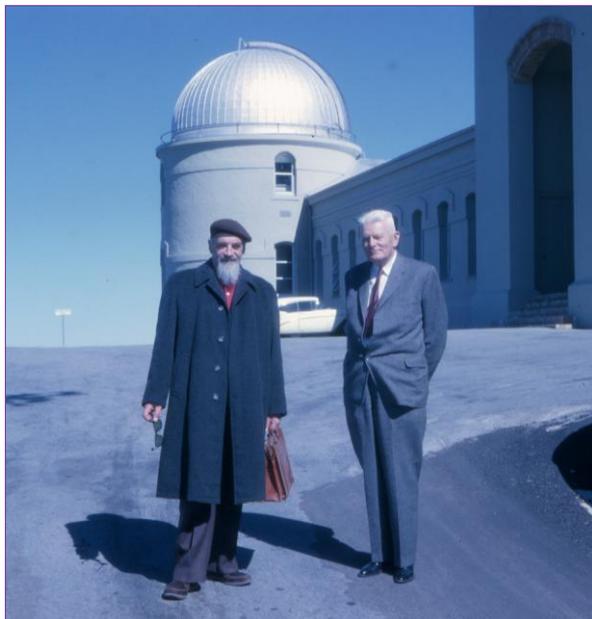

Figure 19: Willem van den Bos and Hamilton Jeffers at Lick Observatory, 1963 (courtesy: James B. Breckinridge).

But how could they publish so much data? It was decided to split it into two catalogues. The *Index Catalogue* (Figure 20, to be known as IDS) was prepared from one 80-column card per star system. It contained positions in 1900 and 2000, discoverer's symbol, discoverer's number, if any, multiplicity, first and last measures with their dates,

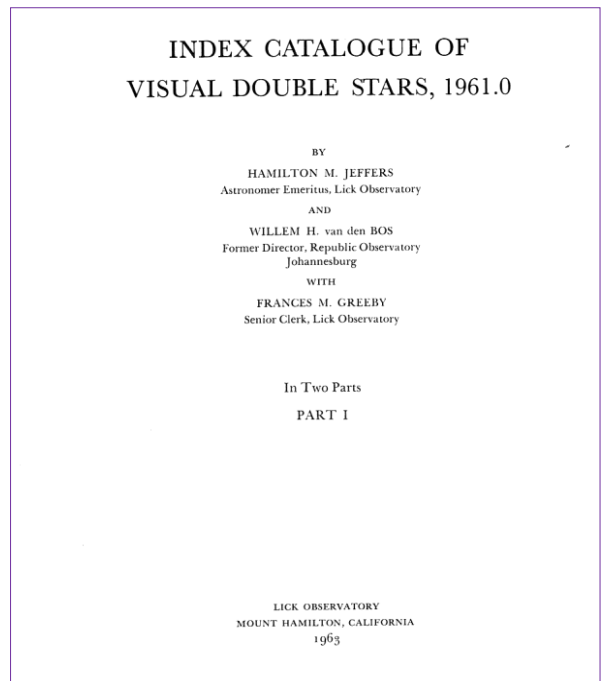

Figure 20: Title page of the IDS (Jeffers, et al., 1963).

number of measures, position angles, separation distances, magnitudes, spectral classes, proper motions, catalogue numbers in previous catalogues, and indication if an orbit had been computed. It was published on paper, as a volume of the *Publications of the Lick Observatory* (Jeffers, et al., 1963). It contained 64,247 pairs and claimed to be "a list of essentially all visual double stars for which measures have been published up to the end of the year 1960" (ibid.: vii). All of the other observations would remain on punched cards. Jeffers thought that three sets of cards would be sufficient for the world's astronomers, one each at Lick and Union Observatories, and the third somewhere in Europe. As they described the project,

… it seems reasonable to suppose that, broadly speaking, double star discovery has now been accomplished. Our principal requirements at present are not further additions to the large number of known double stars, but more and more reliable data on the latter—on their motions as well as their general physical characteristics (van den Bos and Jeffers, 1957: 323).

There was a problem, however. Jeffers turned seventy the year the *Index Catalogue* was published. He had formally retired two years earlier, but had been hired back part-time to see the catalogue through publication. He desperately wanted the Lick Observatory to hire another 'double star man' to replace him and maintain the tradition of Burnham, Aitken, and himself. He wrote letters for several years campaigning for such a hire, but he got nowhere with the new Lick Director, Albert Edward Whitford (Figure 21). Lick had acquired a 120-in (3-meter) reflector, the second largest telescope in the world, in 1958, and Whitford, a pioneer of photoelectric photometry (Tenn, 2013c), was not going to hire an astrometrist to do research with the nineteenth-century Lick refractor. Astrophysics was in vogue and would remain so. Jeffers tried to get a lower level hire: his assistant, James B. Gibson, could maintain the double star program and card catalogue under minimal supervision. Whitford was not interested, and Gibson left to work for Kaj A. Strand (Figure 22) of the United States Naval Observatory (USNO) at its new station near Flagstaff, Arizona.

By 1964 Jeffers was desperate. He suggested that Mrs. Greeby could maintain the card catalogue in one afternoon per week. He wrote emotional letters trying for any solution other than the one he feared: that the catalogues would come under the control of Strand. He wrote Whitford, "Strand is energetic and ambitious, but he is hardly popular as an administrator or as a co-worker." (Jeffers, 1964a). He suggested that if the catalogues could not remain at Lick, then they should go to Paris or Greenwich, where they could be maintained by Paul Muller (1910–2000) or Richard van der Riet Woolley (1906–1986). When Whitford accused him of basing his position on "… poorly concealed reasons of personal dislike …" (Whitford 1964), Jeffers claimed to personally like Strand but,

> On the official or professsional side, though, I have not appreciated his ten years of politicking and influence wielding with the object of getting control of the double star catalogues. However, I cannot help but admire the manner in which his campaign was brought to an apparently successful conclusion. (Jeffers, 1964b).

From most accounts Strand was a difficult person to work with—one person called him "imperious"—but he was certainly an able scientist who got things done. A Dane, he had been an assist-

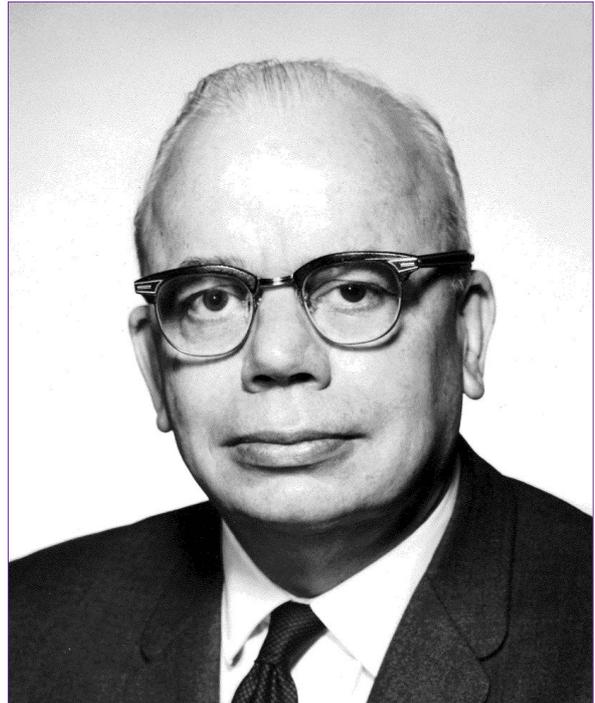

Figure 21: Albert Edward Whitford (1905–2002) in 1968 (courtesy: William Whitford).

ant to Ejnar Hertzsprung at Leiden before emigrating to the United States in 1938. His oral history interviews (Strand, 1983) reveal that he did not lack ego. In any case, Jeffers' efforts to keep the catalogues away from the USNO were all in vain. The International Astronomical Union (IAU) Commission 26 Double and Multiple Stars had appointed a subcommittee on 'Disposition and Management of the Double Star Catalogues'. This group met just before the 1964 IAU meeting in Hamburg and recommended that the catalogues be transferred to the USNO under the following conditions:

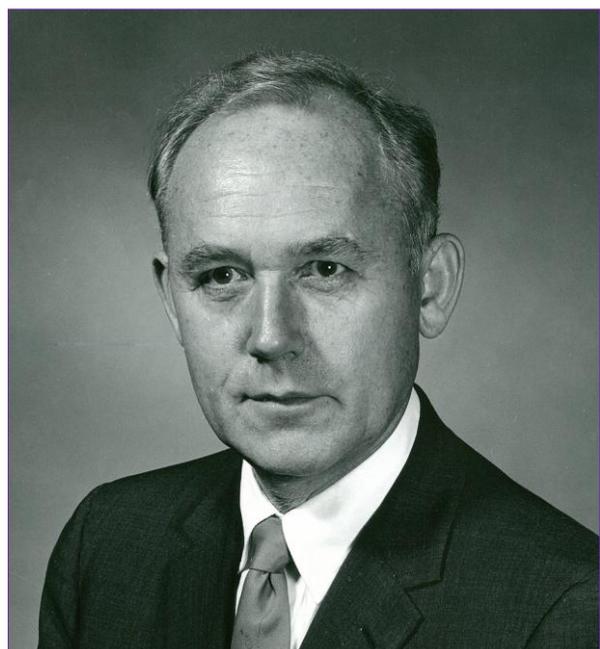

Figure 22: Kaj Aage Strand (1907–2000) (courtesy: U.S. Naval Observatory).

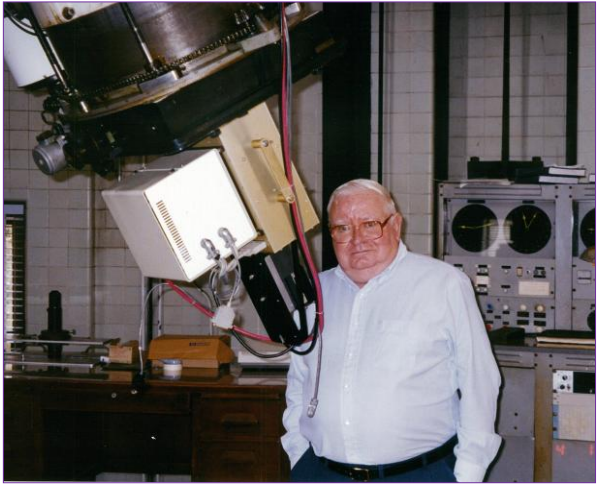

Figure 23: Charles Edmund Worley (1935–1997) (courtesy: U.S. Naval Observatory).

The Double Star Centre at the United States Naval Observatory has the following obligations:
(1) The sole responsibility to maintain the catalogues to date.
(2) To furnish the other centres and the Lick Observatory with supplements when new data become available.
(3) To furnish individuals with data from the two catalogues at cost.

Strand was a member of the subcommittee, but he did not attend the meeting. Whitford, who apparently had asked the Commission what to do with the records, did attend. Jeffers, now fully retired, did not go to Hamburg. The proposal was adopted by the Commission, and Strand became its President (van de Kamp, 1966).

## 6 THE USNO AND THE WDS

Jeffers, who afterward referred to "Whitford's sell-out of the card catalogues," (Jeffers, 1965) may not have been aware that Strand had been promoted in 1963 from Head of the Astrometry and Astrophysics Division to Scientific Director at the USNO. He would never be in direct charge of the double star catalogues. Instead they came under the management of Charles Worley (Figure 23), who had worked at Lick from 1959 to 1961 and who

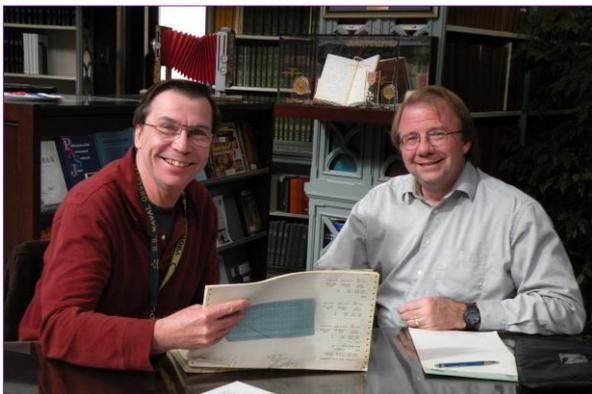

Figure 24: William I. Hartkopf (b. 1951) and Brian D. Mason (b. 1961) in 2012 (photograph by the author).

was respected and liked by his colleagues (Mason, et al., 2007). Worley flew to California in 1965 and took the catalogues back to Washington. They had been converted from cards to magnetic tape, as the cards were too bulky to transport. At the USNO they were transferred back to punched cards. When asked why, Worley replied:

The reason is simple: one needs to do interleaving. That is, to add new observations and even old observations to the data file. And you can't do that easily on tape. The technology of the time, in the 1960s, cards were the best way to do that. We could prepare new cards and insert them in proper spots in the data file. (Worley, 1988).

Worley maintained the catalogues until his death in 1997, just two days before he was scheduled to retire. During his thirty-two years he added 290,400 observational records to the 179,000 he received, and the number of multiple star systems increased from 64,000 to more than 81,000 (Mason, et al., 2007: 30). Alone or with others he produced three *Catalogs of Orbits of Visual Binary Stars* (Worley, 1963; Finsen and Worley, 1970; Worley and Heintz, 1985).

Worley also played a key role in bringing binary star observations—considered by some to be old-fashioned astronomy—into the modern age by obtaining a speckle interferometer for the USNO in 1990 (Douglass, et al., 1997). Mounted on the historic 26-in refractor, which was built in 1873 and moved to its present location in 1893, it is still used regularly for precise measurements of visual binary stars, even though it is in the midst of the brightly-lit capital city (Hartkopf and Mason, 2005). Worley started the USNO practice of hiring astronomers who had worked with Harold McAlister (b. 1949), a leader in developing the speckle technique, at Georgia State University (GSU). Among these are Brian Mason, who earned his Ph.D. at GSU and has directed the Washington Double Star Catalog since Worley's death, and William Hartkopf, who formerly served as Assistant Director of GSU's Center for High Angular Resolution Astronomy. Now Mason and Hartkopf (Figure 24) are the 'keepers of the double stars'.

Today the Washington Double Star Catalog (WDS) is entirely online, and it is updated nightly. A direct descendant of Burnham's BDS, Aitken's ADS, Innes's SDS, and Jeffers and van den Bos's IDS, it is, as its website (WDS, n.d.) states, "… the world's principal database of astrometric double and multiple star information." As of 26 February 2013, it contained 1,201,492 mean measurements of 125,273 star systems.

## 7 NOTES

1. This is oversimplified, as the inclination of the orbital plane to the plane perpendicular to the line of sight, which can also be determined from the visual observations, has to be taken into account in order to compute the actual masses. The determination of masses became much more feasible with the discovery of spectroscopic binaries with their much shorter periods, especially those systems that display eclipses, for which the orbital inclination can also be determined. This paper deals only with visual binaries, systems in which the two stars can be seen or imaged distinctly. Radial velocities of visual binaries can often be used to determine the absolute size of the orbit for those systems that lack a trigonometrical parallax.

2. The Dearborn Observatory began operation in 1865 under the Old University of Chicago (not to be confused with the present one) and after that institution's bankruptcy in 1881 passed to the Chicago Astronomical Society and later to Northwestern University. From 1862 to 1869 the 18.5-in telescope was the largest refractor in the world.

3. Miss Bruce is remembered today for endowing the eponymous medal that has been awarded by the Astronomical Society of the Pacific to the world's leading astronomers since 1898 (Tenn, 1986).

## 8 ACKNOWLEDGEMENTS

A preliminary and much shorter version of this paper was presented to the Historical Astronomy Division of the American Astronomical Society in January 2013. The reader interested in learning more about the early history of double star astronomy may wish to consult the series of papers by T. Lewis (1908). The author is grateful to Alan H. Batten, William I. Hartkopf, Brian D. Mason, and Elizabeth Roemer for advice and comments, and to those who provided images and materials, including James B. Breckinridge, Rebecca Morin, Kelly Jensen, Alan H. Batten, William I. Hartkopf, Eric Shelton, and Ilhan Citak. Excerpts from letters in the Mary Lea Shane Archives of Lick Observatory are published courtesy of Special Collections & Archives, University of California Santa Cruz. This research has made extensive use of the SAO/NASA Astrophysics Data System.

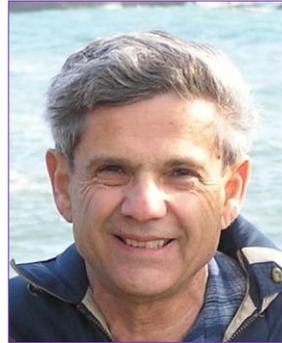

Joseph S. Tenn taught physics and astronomy at Sonoma State University in the California wine country from 1970 to 2009. He now serves as Secretary-Treasurer of the Historical Astronomy Division of the American Astronomical Society and as an Associate Editor of the *JAHH*. He maintains the Bruce Medalists website at http://phys-astro.sonoma.edu/brucemedalists/.